\documentclass[%
 reprint,
 amsmath,amssymb,
 aps,
]{revtex4-2}
\usepackage{graphicx}
\usepackage{float} 
\usepackage{hyperref}
\usepackage{natbib}
\usepackage{xcolor} 
\usepackage{ulem}
\usepackage{dcolumn}
\usepackage{bm}

\begin{document}

\preprint{APS/123-QED}

\title{Half state at $\nu_{tot}$ = -1/2 and its transition in Decoupled Twisted Double Bilayer Graphene}

\author{Ning Ma$^{1}$}
\author{Kenji Watanabe$^{2}$}
\author{Takashi Taniguchi$^{3}$}
\author{Mitali Banerjee$^{1,4}$}
\email{mitali.banerjee@epfl.ch}

\affiliation{$^1$ Institute of Physics, Ecole Polytechnique Fédérale de Lausanne (EPFL), CH-1015 Lausanne, Switzerland}

\affiliation{
$^2$ Research Center for Functional Materials, National Institute for Materials Science, 1-1 Namiki, Tsukuba 305-0044, Japan}

\affiliation{
$^3$ International Center for Materials Nanoarchitectonics, National Institute for Materials Science, 1-1 Namiki, Tsukuba 305-0044, Japan}

\affiliation{
$^4$ Center for Quantum Science and Engineering (QSE Center), École Polytechnique Fédérale de Lausanne (EPFL), 1015, Lausanne, Switzerland}

\date{\today}

\begin{abstract}
\noindent 
The origin of the fractional state at $\nu$ = 1/2 observed in double-layer quantum Hall systems has been under debate for decades. Because of the variation of bilayer charge distribution and interlayer tunneling strength, the half-filling state can be attributed to a two-component(2C) or a one-component(1C) origin, which corresponds to Halperin state and Pffafian state, respectively. Here we report the magnetotransport measurement in decoupled twisted double bilayer graphene(TDBG), which has been proved to be a promising platform for double quantum Hall system. Fractional quantum hall states in both odd and even denominator fillings are observed. We also found that the half-filling state occurs at zero displacement field at $\nu_{tot}$ = -1/2, which is theoretically consistent with two-component Halperin-Laughlin ($\Psi_{331}$) state. Moreover, we report the transition from two-component state at zero $D$ field to one-component non-Abelian state by tunning displacement field. Our observation of the half filling state and its transition from 2C to 1C state provides the tunability of decoupled twisted double bilayer graphene and shed light on the understanding of the ground states at half-filling factor in the double quantum Hall system. 
\end{abstract}

\maketitle


\noindent 
\centerline{\textbf{I. Introduction}}

For decades, novel quantum Hall phenomena including superfluid of exciton condense, even-denominator fractional states,i.e., at 1/2 and 1/4 filling, have been broadly explored in
double-layer quantum Hall systems both in theory\cite{scarola_phase_2001,wen_neutral_1992,alicea_interlayer_2009,barkeshli_non-abelian_2010,geraedts_competing_2015,hou_non-abelian_2025,zhu_fractional_2016} and experiment\cite{halperin_theory_nodate,suen_observation_1992,eisenstein_new_1992,suen_origin_1994}. The even-denominator fractional state at $\nu$ = 1/2 is of particular interest as a possible host of non-Abelian topological orders. In recent years, the nature of even-denominator states in double quantum Hall system has been still under debate\cite{suen_origin_1994,eisenstein_new_1992,liu_fractional_2014,zhu_fractional_2016}, due to the variation of bilayer charge distribution and interlayer tunneling strength between layers among different samples, and experimental study has been limited in two-dimensional electron(hole) systems. 

In a double-layer quantum Hall system, the strong interlayer Coulomb interaction starts to dominate and compete with the intralayer interaction when reducing the seperation between the double layers. The former is characterized by the interlayer distance $d$,  and the latter can be characterized by the magnetic length $l_{B}$ = $\sqrt{\hbar/eB}$, where $\hbar$ is the reduced Planck constant, $e$ is the element charge of electrons and $B$ is the magnetic field. In order to reach the strong coupling regime, an extremely small $d$ is needed. Several previous experiments\cite{li_excitonic_2017,li_excitonic_2017,liu_interlayer_2019,li_pairing_2019,zhang_excitons_2025,nguyen_bilayer_2025,han_anyon_2025} have realized the strongly coupled double-layer system by separating graphene layers with dielectric layers such as hBN of atomic-layer thickness. However, it is still challenging to achieve extremely strong coupled regimes by reducing the thickness of hBN without the inevitable interlayer tunneling. An alternative candidate is the large-angle twisted graphene\cite{rickhaus_correlated_2021,li_strongly_2024,kim_observation_2025}, in which the double layers are stacked close to each other with a distance of sub-nanometer. The large twist-angle guarantees the mismatch of Dirac cones in the reciprocal space and makes the interlayer tunneling negligible.

In this letter, we performed the experiment in large-angle twisted double bilayer graphene, in which each layer is composed of Bernal-stacked bilayer graphene(BLG). Compared with monolayer graphene, bilayer graphene itself provides an additional degree of freedom, orbital number $N$, and can host multiple even-denominator fractional states, which are believed to be non-Abelian states\cite{apalkov_stable_2011,papic_topological_2014,ki_observation_2014,li_even-denominator_2017,kumar_quarter-_2025}. Due to the flexible tuability of graphene by transverse electric field and magnetic field, twisted double bilayer graphene provides a promising platform for studying fractional states, especially at even-denominator fillings.

\begin{figure*}
      \includegraphics[width=\textwidth]{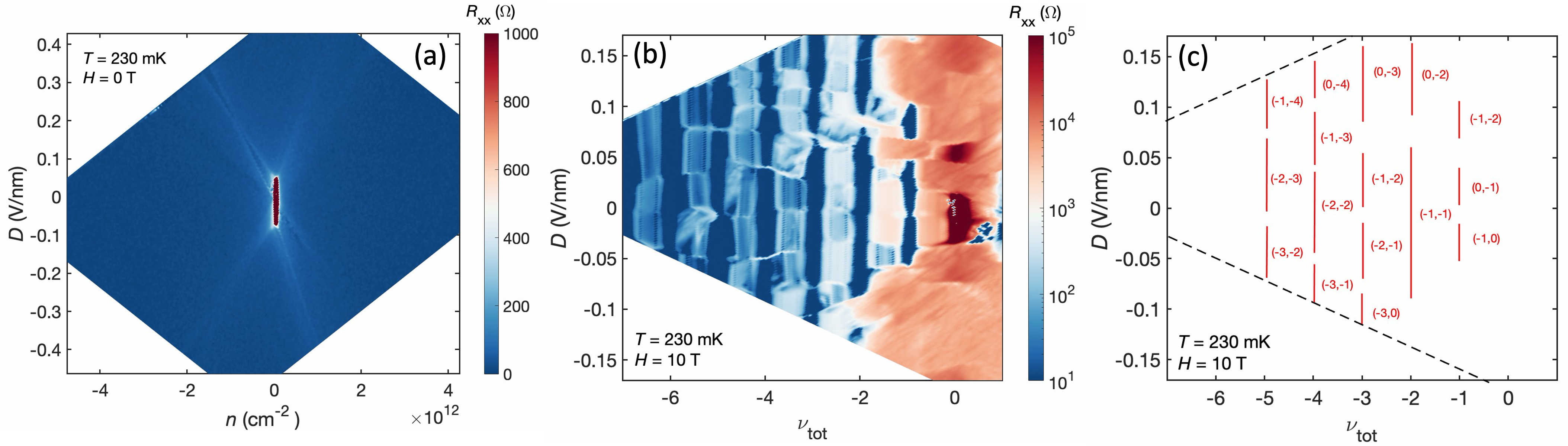}
      \caption{Electrical measurement of 8° twisted double bilayer graphene device. (a) The longitudinal resistance of the measured device as a function of both total carrier density $n$ and displacement field $D$ at 230 mK. The typical dual gate map of large-angle twisted double bilayer graphene devices is consistent with the previous results in the literature. Highly resistive state along the vertical line at zero carrier density is attributed to the gapped states of both layers of bilayer graphene. The full map is divided into several regions by lines indicating the resistive states of bilayer graphene. (b,c) The longitudinal resistance of the measured device on hole-doped side, as a function of both total filling factor $\nu_{tot}$ and displacement field $D$ at 230 mK and 10 T, and the schematic diagram with filling factors in each layer of bilayer graphene labeled, respectively. Both integer and fractional Hall states emerge when applying out-of-plane magnetic field of 10 Tesla. The integer filling factors of both top and bottom layers of BLG are labeled in Fig.1(c) as indicated by the red dashed lines. Transitions between different combinations of filling factors are observed when applying finite displacement field while keeping the total filling factor fixed. The boundaries of each phase can be identified by the high resistance states corresponding to the Landau level crossings. Except the integer states, multiple fractional states of both odd and even denominators can be found in Fig.1(b).
}
      
\end{figure*}

Multiple fractional states are observed on the hole-doped side at both odd and even denominator fillings, in particular, the fractional state at $\nu_{tot}$ = -1/2 at zero displacement field, which can be explained as an interlayer coherent Halperin (331) state\cite{halperin_theory_nodate}. Furthermore, we studied the nature of this half filling state by both displacement field and temperature dependent measurements, and found a transition from two-component state to one-component state by changing displacement field. Our measurement provides an alternative path for controlling the transitions between two-component and one-component half states, compared with previous results in the literature. On the basis of the high tunability of graphene devices, our results suggest decoupled twisted graphene system as a potential platform to study the ground states at half-filling factor in the double quantum Hall system.

\vspace{1em}
\centerline{\textbf{II. Characterization}}
The large-angle twisted double bilayer graphene heterostructure (as shown in Supplementary Materials) is fabricated using the standard cut and stack technique\cite{saito_independent_2020}. A large flake of bilayer graphene is cut into two pieces using AFM tips, then picked up one by one and stacked with a certain twist angle. The device mentioned in the main text has a twist angle of around 8 degrees. The twisted double graphene layers are encapsulated by hBN flakes as dielectric layers.

The longitudinal resistance as a function of both total carrier density $n$ and displacement field $D$ is depicted in Fig.1(a) at 230 mK and 0 T. Here $n$ and $D$ are defined by $n$ = $(C_{tg} V_{tg}+C_{bg} V_{bg})/e-n_{0}$ and $D$ = ($C_{tg} V_{tg}-C_{bg} V_{bg})/2\epsilon_{0}-D_{0}$, respectively. $C_{tg}$ and $C_{bg}$ are the capacitances per area of the top and bottom gate, respectively. $n_{0}$ and $D_{0}$ are the offsets from the residual carrier density and displacement field, respectively. The $n$-$D$ map in the absence of a magnetic field is in good agreement with previous results on large-angle twisted double bilayer graphene\cite{rickhaus_correlated_2021,kim_orbitally_2023,li_strongly_2024}. A high resistive state occurs near zero $n$ and $D$, which is attributed to the crystal fields that stem from the imbalance of the electron distribution between the outer layers and inner layers\cite{rickhaus_gap_2019}. Except for the high resistance region, the map is divided into four regions on the basis of different types of carrier occupied in both top and bottom layers of graphene. For example, the left (right) part of the map indicates that both layers are hole (electron) occupied, whereas the top (bottom) region corresponds to the situation of opposite carrier types in the double layers. In addition, either top or bottom layer of graphene is gapped along the two resistive traces in the map.

We also measured the longitudinal resistance on the hole-doped side in the presence of a magnetic field of 10 T, as shown in Fig.1(b). A global gate of -77 V is applied on the silicon gate in order to observe well-quantized integer and fractional Hall states. The integer filling factors of both top and bottom layers of graphene are labeled in Fig.1(c) according to the positions where Landau level crossings occur. When varying displacement field, transitions between different fractional states are also observed, which will be discussed in detail in the following sections.

\vspace{1em}
\noindent
\centerline{\textbf{III. Jain states at ff = -2 and -3}}

Figure 2 shows the results of longitudinal resistance as a function of both total filling factor (ff) $\nu_{tot}$ and displacement field $D$ measured at a temperature of 230 mK and a perpendicular magnetic field of 10.5 T. The results of Hall resistance can be found in Supplementary Materials. Several fractional states are observed as the resistance minima, which are indicated by the red dashed lines in the map. Filling factors of those fractional states near $\nu_{tot}$ = -2 and -3 are determined by the values of the corresponding Hall conductivity. As shown in Fig.2(a), these fractional states follow Jain's sequence $\nu$ = -(n/2n+1), which can be explained by the single-component composite-fermion model. Moreover, a hierarchy of fractional states obeying 4-flux series can also be found when measured at a nonzero displacement field, as illustrated by the red lines in Fig.3. These 2-flux and 4-flux series are in good agreement with the previous results of graphene\cite{li_even-denominator_2017,zibrov_tunable_2017, li_pairing_2019,kumar_quarter-_2025}, also proving the good quality of our device.

\begin{figure}
    \includegraphics[width=\linewidth]{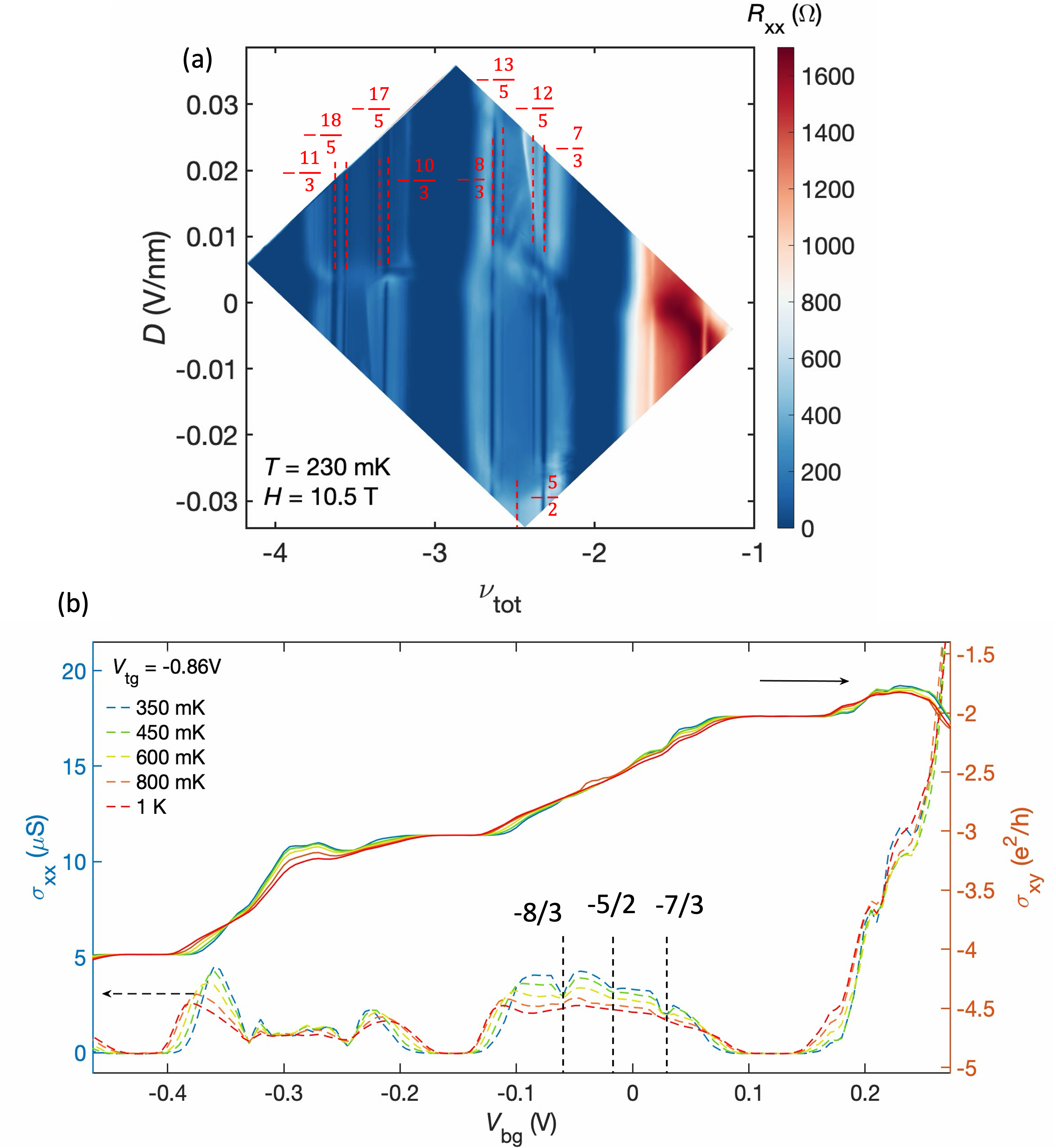}
    \caption{Sequences of fractional states near total filling factor -2 and -3.
(a) The longitudinal resistance near filling factor -2 and -3, as a function of both total carrier density and displacement field at 230 mK and 10.5 T. Fractional states of Jain sequence $\nu$ = -(n/2n+1) emerge when tuning total carrier density at around ff = -2 and -3. The corresponding Hall conductivity maps of these states can be found in Supplementary Materials. In addition to these Jain states, when increasing displacement field, there occurs the transition from Jain states to even denominator states at ff = -2.  As shown in Fig.2(a), Half state of $\nu$ = -5/2 can be found when displacement field $|D|$ is larger than 0.025 V/nm, indicating by the red dashed line in Fig.2(a). (The -5/2 state under positive displacement field can be found in Supplementary Materials).  According to the integer Landau level index in Fig.1(c), the combinations of these fractional states occupied in each layer of BLG can be determined, i.e., the $\nu_{tot}$ = -5/2 state in Fig.2(a) is composed of $\nu$ = -3/2 and $\nu$ = -1 states in top and bottom layers of BLG, respectively. (b) Temperature dependence of both longitudinal and Hall conductivity in the fractional Hall states near total filling factor -2 and -3 are measured from 350 mK to 1 K. According to the temperature dependent measurements, thermal activation gaps of the prominent Jain states and -5/2 states at 10 T are calculated, more details can be found in the Supplementary Materials.
}
\end{figure}

For fractional states that exist under different displacement fields, the combination of filling factors in both top and bottom bilayer graphene can be extracted from the integer index labeled in Fig.1(c), i.e., for the state $\nu_{tot}=-8/3$ marked in Fig.2(a), there undergoes a transition from (-4/3)+(-4/3) at zero displacement field, to (-5/3)+(-1) at finite displacement field. Similar transitions can be found at $\nu_{tot}$ = -18/5, -12/5, -13/5 and -10/3. Interestingly, for $\nu_{tot}$ = -13/5, a resistance dip is also observed at zero displacement field for a balanced population in the bilayer structure. This state is more likely to be attributed to an interlayer coherent state, which is (-1)+(-4/5)+(-4/5). When increasing displacement field, there also occurs the transition from Jain states to even denominator states at ff = -2, i.e., above $\left|D\right|$ = 0.025V/nm, the -5/2 state emerges, as seen in Fig.2(a). According to the integer index in Fig.1(c), the observed -5/2 state is composed of $\nu$ = -3/2 and $\nu$ = -1 in each graphene layer.

\begin{figure}
    \includegraphics[width=1\linewidth]{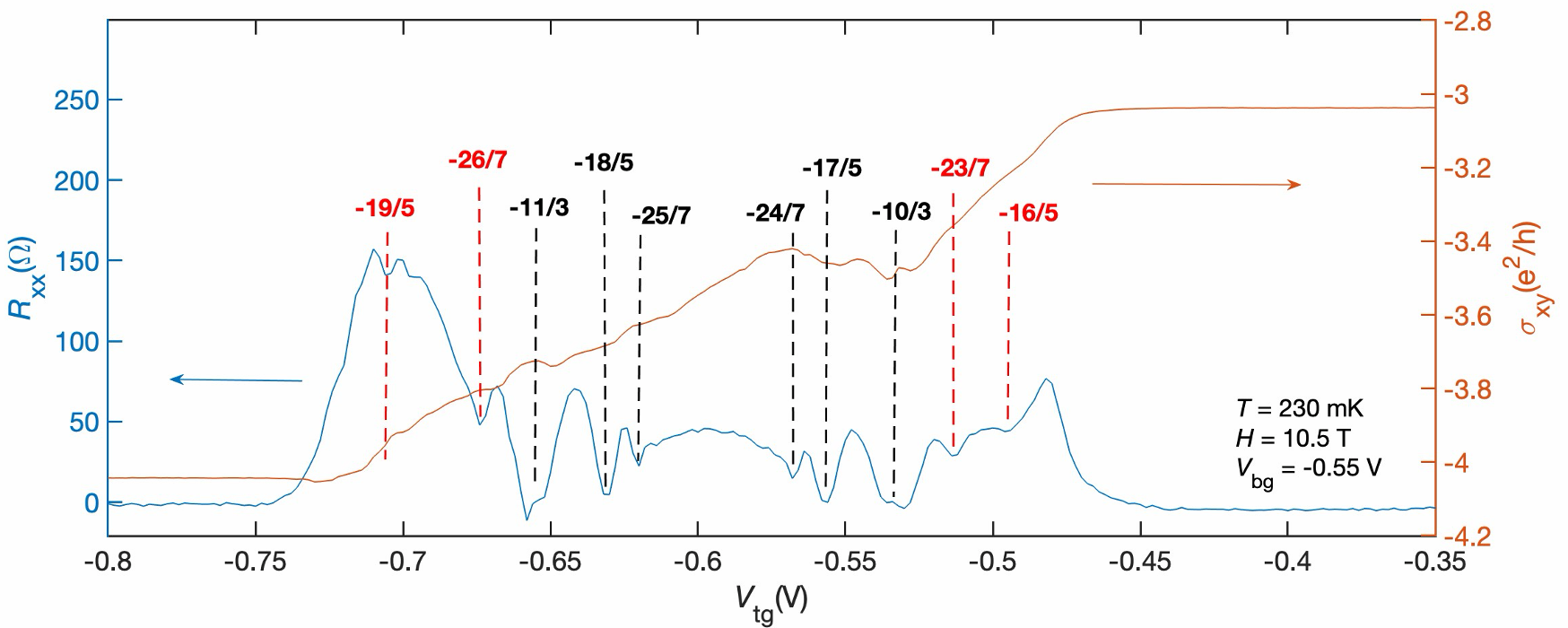}
    \caption{Jain sequences of both the 2-flux and 4-flux composite fermion states. The longitudinal resistance and Hall conductivity are plot as a function of top gate at 230 mK and 10.5 T. Different fractional filling factors corresponding to the 2-flux (4-flux) states are labeled and marked by black (red) dashed lines. All the states are observed under an unzero displacement field.}
\end{figure}

Temperature dependence of the fractional states mentioned above is investigated from 350 mK to 1 K. In Fig.2(b) the results of both longitudinal and Hall conductivity of $\nu$ = -8/3, -7/3 and -5/2 are shown. By performing the temperature dependent measurements, thermal activation gaps at 10 T are extracted from Arrhenius plot. The gaps of -8/3, -7/3 and -5/2 states are 80 mK, 64 mK and 44 mK, respectively. The particle-hole symmetry is indicated around half-filling factor -5/2, according to the identical gaps of $\nu$ = -8/3 and -7/3. More details about the other states near $\nu_{tot}$ = -2 and -3 can be found in the Supplementary Materials.

\vspace{1em}
\noindent
\centerline{\textbf{IV. Half-filling state and its transition at ff = -1}}

Figure 4 shows the emergence of fractional states with odd and even denominators near ff = -1. The red dashed lines in Fig.4(a) indicate the positions of fractional states -1$\pm$1/3, -1$\pm$2/3 and -1/2. In an imbalanced distribution of the charge carrier density, i.e., under nonzero displacement field, the fractional states are composed of integer or fractional states in single layer of BLG, which can be extracted from the index labeled in Fig.1(c) as explained in section III. For a balanced density distribution among the double layers, the fractional states should exist only at even numerator total filling factors, if interlayer Coulomb interactions are not taken into account. The state at $\nu_{tot}$ = -2/3 for zero displacement field is decomposed into a decoupled double layer in which each layer is occupied by -1/3 Laughlin state. According to previous theoretical results, in regimes with  strong interlayer interactions, i.e., the interlayer distance $d$ is much smaller than the magnetic length $l_{B}$, many favorable ground states can be candidates for the -2/3 state, including pseudospin singlet states\cite{faugno_theoretical_2020}, interlayer/intralayer Pfaffian states\cite{barkeshli_non-abelian_2010}, the particle/hole conjugate of 1/3 laughlin states\cite{davenport_spinful_2012}, $Z_{4}$ Read-Rezayi states and composite of 1/3 laughlin states\cite{read_beyond_1999,rezayi_condensation_2010}. A recent study in large-angle twisted bilayer graphene\cite{kim_observation_2025} has shown that the lowest-energy ground state is an interlayer coherent pseudospin singlet state. Similarly, the balanced -4/3 state can be interpreted as an interlayer coherent pseudospin singlet state. 

\begin{figure}
    \includegraphics[width=\linewidth]{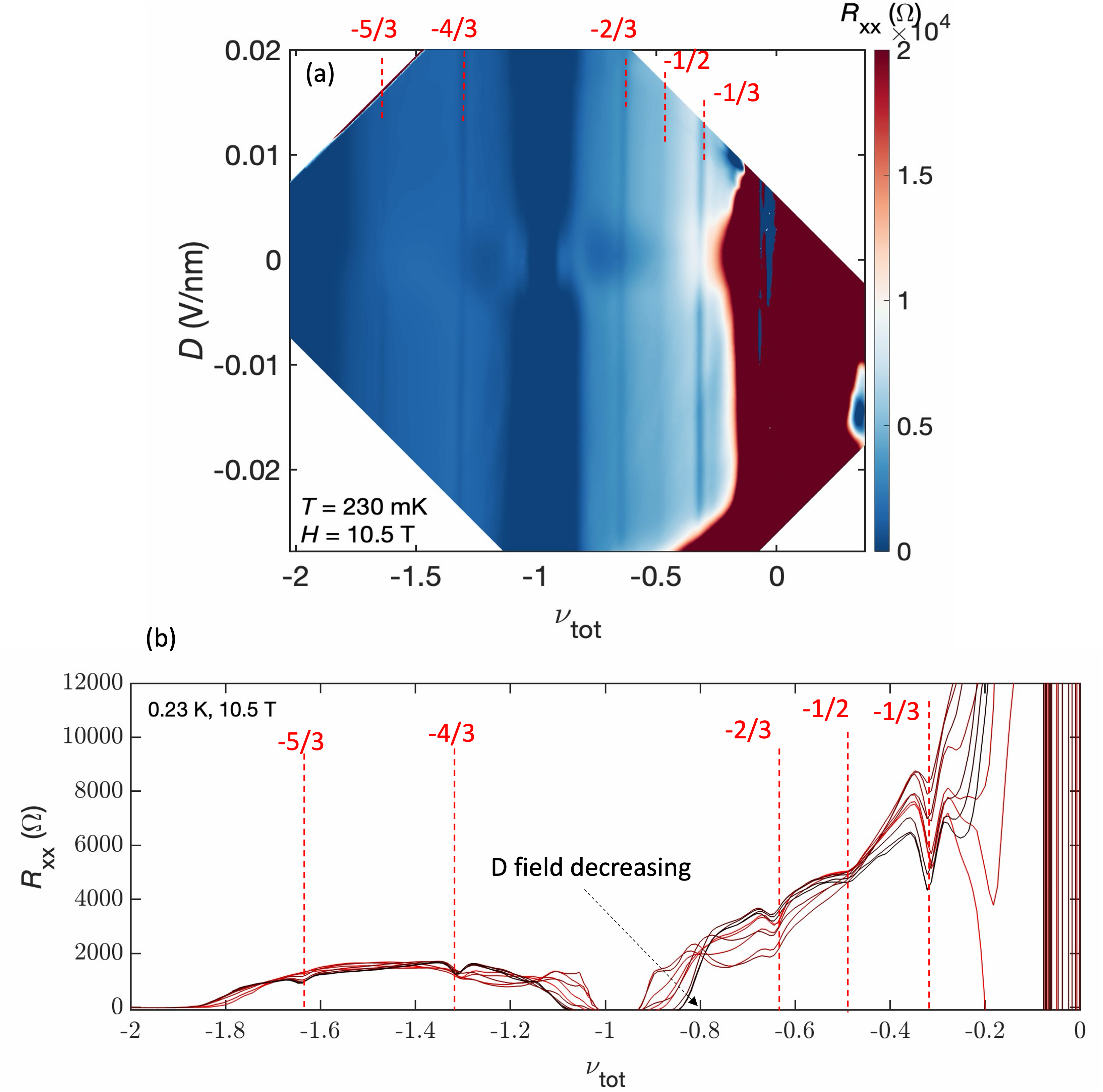}
    \caption{Excitonic states of fractional fillings near $\nu_{tot}$ = -1.
(a) The longitudinal resistance of TDBG as a function of both total filling factors and displacement field at 230 mK and 10.5 T. 1/3 and 2/3 states can be observed on both side of $\nu_{tot}$ = -1, as indicated by the red dashed lines in Fig.4(a). Besides, the -1/2 state is found to coexist with the Jain states even at zero displacement field. Interestingly, the -1/3 state at zero displacement field is attributed to an excitonic ground state which is the fractional analogy to the excitonic states at odd integer filling factors in bilayer quantum Hall systems. Moreover, the -1/2 state is also found to be robust at zero displacement field. In spite of the fact that -1/2 state can be observed at zero displacement field in bilayer graphene, the half state we found in Fig.4(a) is not the same case. Since the displacement field in TDBG dominates the polarization between top and bottom layers, which means at zero D field, there is no specific preference of either top or bottom layer occupation. The -1/2 state observed in Fig.4(a) is more likely to be the ground state of interlayer mixing state. Except that, two triangular-shape regions of resistance dip can be found on both sides of $\nu_{tot}$ = -1, at around zero displacement field.}
\end{figure}

However, resistance minima are also found at $\nu_{tot}$ = -1/3 and -1/2 at zero displacement field. It is shown in Fig.4(a) that these two states can survive in a wide range of displacement field when swept from negative to positive field. According to the previous Monte Carlo simulations\cite{kim_observation_2025}, the ground state of the -1/3 state is an interlayer coherent (333)-state, in which excitons are formed and give rise to incompressible behavior. It can be considered as a fractional analog of the (111)-state\cite{wen_neutral_1992,yang_quantum_1994,moon_spontaneous_1995,li_strongly_2024,zhang_excitons_2025} at $\nu_{tot}$ = -1 due to the formation of excitons made of skyrmions that host meron-antimeron pairs with opposite charges in top and bottom layers. The only difference is that in the case of -1/3 state the excitons are formed by quasiparticles with fractional charges\cite{zhang_excitons_2025,kim_observation_2025}.

\begin{figure}[b]
    \includegraphics[width=0.9\linewidth]{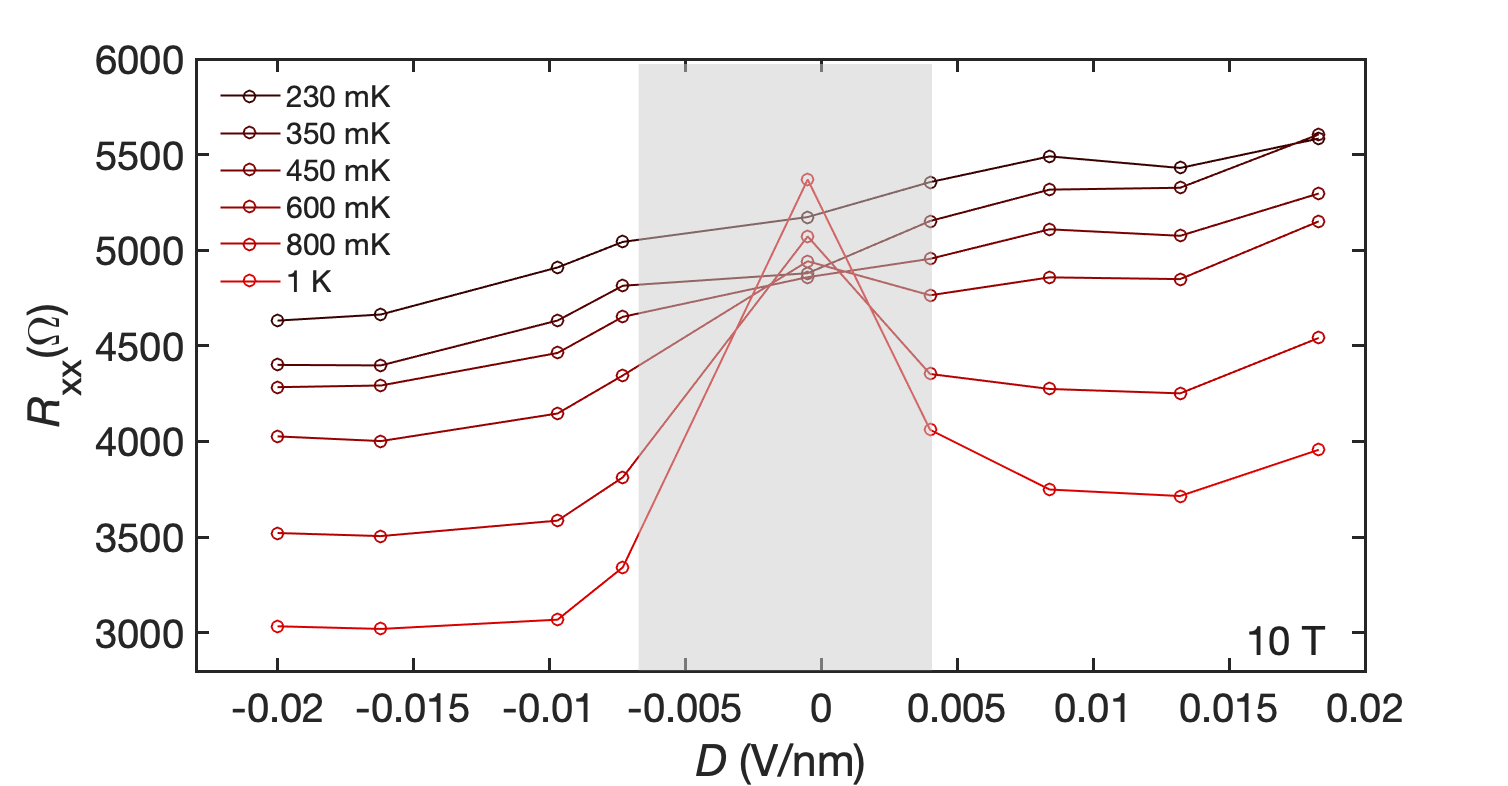}
    \caption{Displacement field and temperature dependence of the longitudinal resistance at $\nu_{tot}$ = -1/2. The $R_{xx}$ as a function of displacement field is measured at different temperatures, with total filling factor fixed at -1/2. At small $D$ field region, denoted by the gray shading in the figure, the resistance follows strong $D$ field-dependence especially at high temperature, in contrast to the rest of the diagram, where the state shows robustness against $D$ field, indicating the robustness against the charge distribution asymmetry. }

\end{figure}

The $\nu_{tot}$ = -1/2 state is reported before in different systems, i.e., double layer two-dimensional electron systems\cite{eisenstein_new_1992,suen_observation_1992}, wide GaAs quantum wells\cite{suen_observation_1992,suen_origin_1994,liu_fractional_2014} and bilayer graphene\cite{ki_observation_2014,li_even-denominator_2017,zibrov_tunable_2017}. The ground state of the $\nu_{tot}$ = -1/2 state, is believed to be either a one-component, non-Abelian, Moore-Read state\cite{moore_nonabelions_1991,greiter_paired_1992,zhu_fractional_2016}, or alternatively, a two-component, Abelian, Halperin-Laughlin (331)-state\cite{halperin_theory_nodate}. In our measurement the -1/2 state at zero displacement field is consistent with the two-component (331) state based on the following discussion. The Two-component state is sensitive to the charge distribution symmetry\cite{suen_origin_1994,liu_fractional_2014,singh_fractional_2025} between the top and bottom layers, which could be modulated by applying a finite displacement field. As shown in Fig.5, longitudinal resistance was measured as a function of displacement field while total filling factor was kept at $\nu_{tot}$ = -1/2. The $R_{xx}$ at finite $D$ field, i.e., outside the gray shading region, shows little dependence of $D$ field, compared to the strong impact in small $D$ field ranges, i.e., the gray shading region. Indeed, when displacement field is applied, charges tend to be polarized to either top or bottom layer, resulting in a breakdown of the interlayer coherent state and the formation of a layer-polarized fractional state\cite{kim_orbitally_2023,kim_observation_2025,kim_correlated_2026}. 

Since $\nu$ = -1/2 state in bilayer graphene hosts the non-Abelian state, based on previous theoretical\cite{apalkov_stable_2011,papic_topological_2014} and experimental results\cite{ki_observation_2014,li_even-denominator_2017,kumar_quarter-_2025}, we claim that a transition from two-component Abelian state to one-component non-Abelian state is observed by tuning displacement field. The temperature dependence of $R_{xx}$ in Fig.5 and Supplementary Materials also indicates that the half filling states at zero and nonzero $D$ field can have different origins. When the temperature increases, the state at $D$ = 0 undergoes a transition from a fractional Hall state at 230 mK with resistance minimum to an insulating state with resistance maximum at 1 K, corresponding to the normal Landau level crossing. It reveals the competition between thermal activation and the interaction-induced gap. At high temperature, the two-component fractional state diminishes and the system turns out to host integer Hall states with Landau level crossings. In contrast, the half state at non-zero $D$ field can survive at higher temperature, consistent with previously calculated large energy gaps\cite{assouline_energy_2024}.

\vspace{1em}
\noindent
\centerline{\textbf{V. Discussion and conclusions}}

In conclusion, we have studied the fractional Hall states in decoupled twisted double bilayer graphene. Both odd and even denominator states are observed at multiple filling factors on the hole side. We analyze the layer-dependent filling factor combinations for different fractional states, and activation gaps of those fractional states are extracted from the Arrhenius analysis. Besides, we discuss the transition of those fractional states from zero $D$ field to finite $D$ field, especially the case of half filling state at $\nu_{tot}$ = -1/2, which unveils a transition from Abelian Halperin state to non-Abelian Pfaffian state. It is worth pointing out that the transition between one-component state and two-component state has been reported in GaAs quantum wells by multiple approaches, i.e., in-plane magnetic field\cite{lay_one-component_1997} and tuning carrier densities\cite{singh_fractional_2025}. Compared with these previous results, our observation sheds light on the displacement field driven transition between non-Abelian state and Abelian state, and proves decoupled twisted double bilayer graphene as a candidate platform for investigating fractional coherent states and for constructing non-Abelian coherent states\cite{hou_non-abelian_2025} in the future.







\newpage

\bibliography{references}

@misc{rezayi_condensation_2010,
    title = {Condensation of fractional excitons, non-{Abelian} states in double-layer quantum {Hall} systems and {$Z_4$} parafermions},
    url = {http://arxiv.org/abs/1007.2022},
    doi = {10.48550/arXiv.1007.2022},
    urldate = {2026-04-26},
    publisher = {arXiv},
    author = {Rezayi, Edward and Wen, Xiao-Gang and Read, N.},
    month = jul,
    year = {2010},
    note = {arXiv:1007.2022 [cond-mat]},
    
}

@article{saito_independent_2020,
    title = {Independent superconductors and correlated insulators in twisted bilayer graphene},
    volume = {16},
    copyright = {2020 The Author(s), under exclusive licence to Springer Nature Limited},
    issn = {1745-2481},
    url = {https://www.nature.com/articles/s41567-020-0928-3},
    doi = {10.1038/s41567-020-0928-3},
    number = {9},
    urldate = {2025-07-30},
    journal = {Nature Physics},
    publisher = {Nature Publishing Group},
    author = {Saito, Yu and Ge, Jingyuan and Watanabe, Kenji and Taniguchi, Takashi and Young, Andrea F.},
    month = sep,
    year = {2020},
    pages = {926--930},
}

@article{faugno_theoretical_2020,
    title = {Theoretical phase diagram of two-component composite fermions in double-layer graphene},
    volume = {101},
    issn = {2469-9950, 2469-9969},
    url = {https://link.aps.org/doi/10.1103/PhysRevB.101.085412},
    doi = {10.1103/PhysRevB.101.085412},
    number = {8},
    urldate = {2026-04-26},
    journal = {Physical Review B},
    author = {Faugno, W. N. and Balram, Ajit C. and Wójs, A. and Jain, J. K.},
    month = feb,
    year = {2020},
    pages = {085412},
}

@article{barkeshli_non-abelian_2010,
    title = {Non-{Abelian} two-component fractional quantum {Hall} states},
    volume = {82},
    copyright = {http://link.aps.org/licenses/aps-default-license},
    issn = {1098-0121, 1550-235X},
    url = {https://link.aps.org/doi/10.1103/PhysRevB.82.233301},
    doi = {10.1103/PhysRevB.82.233301},
    number = {23},
    urldate = {2026-04-26},
    journal = {Physical Review B},
    author = {Barkeshli, Maissam and Wen, Xiao-Gang},
    month = dec,
    year = {2010},
    pages = {233301},
}

@article{davenport_spinful_2012,
    title = {Spinful composite fermions in a negative effective field},
    volume = {85},
    url = {https://link.aps.org/doi/10.1103/PhysRevB.85.245303},
    doi = {10.1103/PhysRevB.85.245303},
    number = {24},
    urldate = {2026-04-26},
    journal = {Physical Review B},
    publisher = {American Physical Society},
    author = {Davenport, Simon C. and Simon, Steven H.},
    month = jun,
    year = {2012},
    pages = {245303},
}

@article{read_beyond_1999,
    title = {Beyond paired quantum {Hall} states: {Parafermions} and incompressible states in the first excited {Landau} level},
    volume = {59},
    copyright = {http://link.aps.org/licenses/aps-default-license},
    issn = {0163-1829, 1095-3795},
    shorttitle = {Beyond paired quantum {Hall} states},
    url = {https://link.aps.org/doi/10.1103/PhysRevB.59.8084},
    doi = {10.1103/PhysRevB.59.8084},
    number = {12},
    urldate = {2026-04-26},
    journal = {Physical Review B},
    author = {Read, N. and Rezayi, E.},
    month = mar,
    year = {1999},
    pages = {8084--8092},
}

@article{kim_observation_2025,
    title = {Observation of 1/3 fractional quantum {Hall} physics in balanced large angle twisted bilayer graphene},
    volume = {16},
    copyright = {2024 The Author(s)},
    issn = {2041-1723},
    url = {https://www.nature.com/articles/s41467-024-55486-2},
    doi = {10.1038/s41467-024-55486-2},
    number = {1},
    urldate = {2025-07-19},
    journal = {Nature Communications},
    publisher = {Nature Publishing Group},
    author = {Kim, Dohun and Jin, Seyoung and Taniguchi, Takashi and Watanabe, Kenji and Smet, Jurgen H. and Cho, Gil Young and Kim, Youngwook},
    month = jan,
    year = {2025},
    pages = {179},
}

@article{li_strongly_2024,
    title = {Strongly coupled magneto-exciton condensates in large-angle twisted double bilayer graphene},
    volume = {15},
    copyright = {2024 The Author(s)},
    issn = {2041-1723},
    url = {https://www.nature.com/articles/s41467-024-49406-7},
    doi = {10.1038/s41467-024-49406-7},
    number = {1},
    urldate = {2025-07-19},
    journal = {Nature Communications},
    publisher = {Nature Publishing Group},
    author = {Li, Qingxin and Chen, Yiwei and Wei, LingNan and Chen, Hong and Huang, Yan and Zhu, Yujian and Zhu, Wang and An, Dongdong and Song, Junwei and Gan, Qikang and Zhang, Qi and Watanabe, Kenji and Taniguchi, Takashi and Shi, Xiaoyang and Novoselov, Kostya S. and Wang, Rui and Yu, Geliang and Wang, Lei},
    month = jun,
    year = {2024},
    pages = {5065},
}

@article{yang_quantum_1994,
    title = {Quantum ferromagnetism and phase transitions in double-layer quantum {Hall} systems},
    volume = {72},
    copyright = {http://link.aps.org/licenses/aps-default-license},
    issn = {0031-9007},
    url = {https://link.aps.org/doi/10.1103/PhysRevLett.72.732},
    doi = {10.1103/PhysRevLett.72.732},
    number = {5},
    urldate = {2025-11-01},
    journal = {Physical Review Letters},
    author = {Yang, Kun and Moon, K. and Zheng, L. and MacDonald, A. H. and Girvin, S. M. and Yoshioka, D. and Zhang, Shou-Cheng},
    month = jan,
    year = {1994},
    pages = {732--735},
}

@article{moon_spontaneous_1995,
    title = {Spontaneous interlayer coherence in double-layer quantum {Hall} systems: {Charged} vortices and {Kosterlitz}-{Thouless} phase transitions},
    volume = {51},
    copyright = {http://link.aps.org/licenses/aps-default-license},
    issn = {0163-1829, 1095-3795},
    shorttitle = {Spontaneous interlayer coherence in double-layer quantum {Hall} systems},
    url = {https://link.aps.org/doi/10.1103/PhysRevB.51.5138},
    doi = {10.1103/PhysRevB.51.5138},
    number = {8},
    urldate = {2025-10-30},
    journal = {Physical Review B},
    author = {Moon, K. and Mori, H. and Yang, Kun and Girvin, S. M. and MacDonald, A. H. and Zheng, L. and Yoshioka, D. and Zhang, Shou-Cheng},
    month = feb,
    year = {1995},
    pages = {5138--5170},
}

@article{wen_neutral_1992,
    title = {Neutral superfluid modes and ‘‘magnetic’’ monopoles in multilayered quantum {Hall} systems},
    volume = {69},
    copyright = {http://link.aps.org/licenses/aps-default-license},
    issn = {0031-9007},
    url = {https://link.aps.org/doi/10.1103/PhysRevLett.69.1811},
    doi = {10.1103/PhysRevLett.69.1811},
    number = {12},
    urldate = {2026-04-26},
    journal = {Physical Review Letters},
    author = {Wen, Xiao-Gang and Zee, A.},
    month = sep,
    year = {1992},
    pages = {1811--1814},
}

@article{zhang_excitons_2025,
    title = {Excitons in the fractional quantum {Hall} effect},
    volume = {637},
    copyright = {2025 The Author(s), under exclusive licence to Springer Nature Limited},
    issn = {1476-4687},
    url = {https://www.nature.com/articles/s41586-024-08274-3},
    doi = {10.1038/s41586-024-08274-3},
    number = {8045},
    urldate = {2025-09-30},
    journal = {Nature},
    publisher = {Nature Publishing Group},
    author = {Zhang, Naiyuan J. and Nguyen, Ron Q. and Batra, Navketan and Liu, Xiaoxue and Watanabe, Kenji and Taniguchi, Takashi and Feldman, D. E. and Li, J. I. A.},
    month = jan,
    year = {2025},
    pages = {327--332},
}

@article{eisenstein_new_1992,
    title = {New fractional quantum {Hall} state in double-layer two-dimensional electron systems},
    volume = {68},
    copyright = {http://link.aps.org/licenses/aps-default-license},
    issn = {0031-9007},
    url = {https://link.aps.org/doi/10.1103/PhysRevLett.68.1383},
    doi = {10.1103/PhysRevLett.68.1383},
    number = {9},
    urldate = {2026-03-10},
    journal = {Physical Review Letters},
    author = {Eisenstein, J. P. and Boebinger, G. S. and Pfeiffer, L. N. and West, K. W. and He, Song},
    month = mar,
    year = {1992},
    pages = {1383--1386},
}

@article{suen_observation_1992,
    title = {Observation of a $\nu = 1/2$ fractional quantum {Hall} state in a double-layer electron system},
    volume = {68},
    copyright = {http://link.aps.org/licenses/aps-default-license},
    issn = {0031-9007},
    url = {https://link.aps.org/doi/10.1103/PhysRevLett.68.1379},
    doi = {10.1103/PhysRevLett.68.1379},
    number = {9},
    urldate = {2026-03-10},
    journal = {Physical Review Letters},
    author = {Suen, Y. W. and Engel, L. W. and Santos, M. B. and Shayegan, M. and Tsui, D. C.},
    month = mar,
    year = {1992},
    pages = {1379--1382},
}

@article{liu_fractional_2014,
    title = {Fractional {Quantum} {Hall} {Effect} at $\nu = 1/2$ in {Hole} {Systems} {Confined} to {GaAs} {Quantum} {Wells}},
    volume = {112},
    copyright = {http://link.aps.org/licenses/aps-default-license},
    issn = {0031-9007, 1079-7114},
    url = {https://link.aps.org/doi/10.1103/PhysRevLett.112.046804},
    doi = {10.1103/PhysRevLett.112.046804},
    number = {4},
    urldate = {2026-03-10},
    journal = {Physical Review Letters},
    author = {Liu, Yang and Graninger, A. L. and Hasdemir, S. and Shayegan, M. and Pfeiffer, L. N. and West, K. W. and Baldwin, K. W. and Winkler, R.},
    month = jan,
    year = {2014},
    pages = {046804},
}

@article{suen_origin_1994,
    title = {Origin of the $\nu = 1/2$ fractional quantum {Hall} state in wide single quantum wells},
    volume = {72},
    copyright = {http://link.aps.org/licenses/aps-default-license},
    issn = {0031-9007},
    url = {https://link.aps.org/doi/10.1103/PhysRevLett.72.3405},
    doi = {10.1103/PhysRevLett.72.3405},
    number = {21},
    urldate = {2026-04-26},
    journal = {Physical Review Letters},
    author = {Suen, Y. W. and Manoharan, H. C. and Ying, X. and Santos, M. B. and Shayegan, M.},
    month = may,
    year = {1994},
    pages = {3405--3408},
}

@article{ki_observation_2014,
    title = {Observation of Even Denominator Fractional Quantum Hall Effect in Suspended Bilayer Graphene},
    volume = {14},
    issn = {1530-6984},
    url = {https://doi.org/10.1021/nl5003922},
    doi = {10.1021/nl5003922},
    number = {4},
    urldate = {2026-03-10},
    journal = {Nano Letters},
    publisher = {American Chemical Society},
    author = {Ki, Dong-Keun and Fal’ko, Vladimir I. and Abanin, Dmitry A. and Morpurgo, Alberto F.},
    month = apr,
    year = {2014},
    pages = {2135--2139},
}

@article{li_even-denominator_2017,
    title = {Even-denominator fractional quantum {Hall} states in bilayer graphene},
    volume = {358},
    url = {https://www.science.org/doi/10.1126/science.aao2521},
    doi = {10.1126/science.aao2521},
    number = {6363},
    urldate = {2025-03-16},
    journal = {Science},
    publisher = {American Association for the Advancement of Science},
    author = {Li, J. I. A. and Tan, C. and Chen, S. and Zeng, Y. and Taniguchi, T. and Watanabe, K. and Hone, J. and Dean, C. R.},
    month = nov,
    year = {2017},
    pages = {648--652},
}

@article{zibrov_tunable_2017,
    title = {Tunable interacting composite fermion phases in a half-filled bilayer-graphene {Landau} level},
    volume = {549},
    copyright = {2017 Macmillan Publishers Limited, part of Springer Nature. All rights reserved.},
    issn = {1476-4687},
    url = {https://www.nature.com/articles/nature23893},
    doi = {10.1038/nature23893},
    number = {7672},
    urldate = {2026-01-13},
    journal = {Nature},
    publisher = {Nature Publishing Group},
    author = {Zibrov, A. A. and Kometter, C. and Zhou, H. and Spanton, E. M. and Taniguchi, T. and Watanabe, K. and Zaletel, M. P. and Young, A. F.},
    month = sep,
    year = {2017},
    pages = {360--364},
}

@article{moore_nonabelions_1991,
    title = {Nonabelions in the fractional quantum hall effect},
    volume = {360},
    issn = {0550-3213},
    url = {https://www.sciencedirect.com/science/article/pii/055032139190407O},
    doi = {10.1016/0550-3213(91)90407-O},
    number = {2},
    urldate = {2026-04-26},
    journal = {Nuclear Physics B},
    author = {Moore, Gregory and Read, Nicholas},
    month = aug,
    year = {1991},
    pages = {362--396},
}

@article{greiter_paired_1992,
    title = {Paired {Hall} states in double-layer electron systems},
    volume = {46},
    url = {https://link.aps.org/doi/10.1103/PhysRevB.46.9586},
    doi = {10.1103/PhysRevB.46.9586},
    number = {15},
    urldate = {2026-04-26},
    journal = {Physical Review B},
    publisher = {American Physical Society},
    author = {Greiter, Martin and Wen, X. G. and Wilczek, Frank},
    month = oct,
    year = {1992},
    pages = {9586--9589},
}

@article{zhu_fractional_2016,
    title = {Fractional quantum {Hall} bilayers at half filling: {Tunneling}-driven non-{Abelian} phase},
    volume = {94},
    copyright = {http://link.aps.org/licenses/aps-default-license},
    issn = {2469-9950, 2469-9969},
    shorttitle = {Fractional quantum {Hall} bilayers at half filling},
    url = {https://link.aps.org/doi/10.1103/PhysRevB.94.245147},
    doi = {10.1103/PhysRevB.94.245147},
    number = {24},
    urldate = {2026-04-26},
    journal = {Physical Review B},
    author = {Zhu, W. and Liu, Zhao and Haldane, F. D. M. and Sheng, D. N.},
    month = dec,
    year = {2016},
    pages = {245147},
}

@article{halperin_theory_nodate,
    title = {Theory of the quantized Hall conductance},
    author = {Halperin, B. I.},
    journal = {Helv Phys Acta},
    volume = {56},
    pages = {75--102},
    year = {1983}
}

@article{kumar_quarter-_2025,
    title = {Quarter- and half-filled quantum {Hall} states and their topological orders revealed by daughter states in bilayer graphene},
    volume = {16},
    copyright = {2025 The Author(s)},
    issn = {2041-1723},
    url = {https://www.nature.com/articles/s41467-025-62650-9},
    doi = {10.1038/s41467-025-62650-9},
    number = {1},
    urldate = {2026-04-26},
    journal = {Nature Communications},
    publisher = {Nature Publishing Group},
    author = {Kumar, Ravi and Haug, André and Kim, Jehyun and Yutushui, Misha and Khudiakov, Konstantin and Bhardwaj, Vishal and Ilin, Alexey and Watanabe, K. and Taniguchi, T. and Mross, David F. and Ronen, Yuval},
    month = aug,
    year = {2025},
    pages = {7255},
}

@article{li_pairing_2019,
    title = {Pairing states of composite fermions in double-layer graphene},
    volume = {15},
    copyright = {2019 The Author(s), under exclusive licence to Springer Nature Limited},
    issn = {1745-2481},
    url = {https://www.nature.com/articles/s41567-019-0547-z},
    doi = {10.1038/s41567-019-0547-z},
    number = {9},
    urldate = {2025-10-30},
    journal = {Nature Physics},
    publisher = {Nature Publishing Group},
    author = {Li, J. I. A. and Shi, Q. and Zeng, Y. and Watanabe, K. and Taniguchi, T. and Hone, J. and Dean, C. R.},
    month = sep,
    year = {2019},
    pages = {898--903},
}

@article{singh_fractional_2025,
    title = {Fractional {Quantum} {Hall} {State} at $\nu = 1 / 2$ with {Energy} {Gap} {Up} to 6 {K} and {Possible} {Transition} from the {One}- to {Two}-{Component} {State}},
    volume = {135},
    issn = {0031-9007, 1079-7114},
    url = {https://link.aps.org/doi/10.1103/ywpx-qm7d},
    doi = {10.1103/ywpx-qm7d},
    number = {24},
    urldate = {2026-04-20},
    journal = {Physical Review Letters},
    author = {Singh, Siddharth Kumar and Wang, Chengyu and Gupta, Adbhut and Baldwin, Kirk W. and Pfeiffer, Loren N. and Shayegan, Mansour},
    month = dec,
    year = {2025},
    pages = {246603},
}

@article{kim_orbitally_2023,
    title = {Orbitally {Controlled} {Quantum} {Hall} {States} in {Decoupled} {Two}-{Bilayer} {Graphene} {Sheets}},
    volume = {10},
    copyright = {© 2023 The Authors. Advanced Science published by Wiley-VCH GmbH},
    issn = {2198-3844},
    url = {https://onlinelibrary.wiley.com/doi/abs/10.1002/advs.202300574},
    doi = {10.1002/advs.202300574},
    number = {23},
    urldate = {2025-11-13},
    journal = {Advanced Science},
    author = {Kim, Soyun and Kim, Dohun and Watanabe, Kenji and Taniguchi, Takashi and Smet, Jurgen H. and Kim, Youngwook},
    year = {2023},
    pages = {2300574},
}

@article{kim_correlated_2026,
    title = {Correlated {Interlayer} {Quantum} {Hall} {State} in {Large}-{Angle} {Twisted} {Trilayer} {Graphene}},
    volume = {26},
    issn = {1530-6984},
    url = {https://doi.org/10.1021/acs.nanolett.5c04989},
    doi = {10.1021/acs.nanolett.5c04989},
    number = {1},
    urldate = {2026-04-27},
    journal = {Nano Letters},
    publisher = {American Chemical Society},
    author = {Kim, Dohun and Lee, Gyeoul and Leconte, Nicolas and Jin, Seyoung and Taniguchi, Takashi and Watanabe, Kenji and Jung, Jeil and Cho, Gil Young and Kim, Youngwook},
    month = jan,
    year = {2026},
    pages = {231--237},
}

@article{papic_topological_2014,
    title = {Topological {Phases} in the {Zeroth} {Landau} {Level} of {Bilayer} {Graphene}},
    volume = {112},
    url = {https://link.aps.org/doi/10.1103/PhysRevLett.112.046602},
    doi = {10.1103/PhysRevLett.112.046602},
    number = {4},
    urldate = {2026-04-27},
    journal = {Physical Review Letters},
    publisher = {American Physical Society},
    author = {Papić, Z. and Abanin, D. A.},
    month = jan,
    year = {2014},
    pages = {046602},
}

@article{apalkov_stable_2011,
    title = {Stable {Pfaffian} {State} in {Bilayer} {Graphene}},
    volume = {107},
    url = {https://link.aps.org/doi/10.1103/PhysRevLett.107.186803},
    doi = {10.1103/PhysRevLett.107.186803},
    number = {18},
    urldate = {2026-04-27},
    journal = {Physical Review Letters},
    publisher = {American Physical Society},
    author = {Apalkov, Vadim M. and Chakraborty, Tapash},
    month = oct,
    year = {2011},
    pages = {186803},
}

@article{assouline_energy_2024,
    title = {Energy {Gap} of the {Even}-{Denominator} {Fractional} {Quantum} {Hall} {State} in {Bilayer} {Graphene}},
    volume = {132},
    issn = {0031-9007, 1079-7114},
    url = {https://link.aps.org/doi/10.1103/PhysRevLett.132.046603},
    doi = {10.1103/PhysRevLett.132.046603},
    number = {4},
    urldate = {2025-03-16},
    journal = {Physical Review Letters},
    author = {Assouline, Alexandre and Wang, Taige and Zhou, Haoxin and Cohen, Liam A. and Yang, Fangyuan and Zhang, Ruining and Taniguchi, Takashi and Watanabe, Kenji and Mong, Roger S. K. and Zaletel, Michael P. and Young, Andrea F.},
    month = jan,
    year = {2024},
    pages = {046603},
}

@article{lay_one-component_1997,
    title = {One-component to two-component transition of the $\nu=2/3$ fractional quantum {Hall} effect in a wide quantum well induced by an in-plane magnetic field},
    volume = {56},
    url = {https://link.aps.org/doi/10.1103/PhysRevB.56.R7092},
    doi = {10.1103/PhysRevB.56.R7092},
    number = {12},
    urldate = {2026-04-20},
    journal = {Physical Review B},
    publisher = {American Physical Society},
    author = {Lay, T. S. and Jungwirth, T. and Smrčka, L. and Shayegan, M.},
    month = sep,
    year = {1997},
    pages = {R7092--R7095},
}

@misc{hou_non-abelian_2025,
    title = {Non-{Abelian} interlayer coherent fractional quantum {Hall} states},
    url = {http://arxiv.org/abs/2501.06041},
    doi = {10.48550/arXiv.2501.06041},
    urldate = {2026-04-27},
    publisher = {arXiv},
    author = {Hou, Xiang-Jian and Wang, Lei and Wu, Ying-Hai},
    month = feb,
    year = {2025},
    note = {arXiv:2501.06041 [cond-mat]},
}

@article{rickhaus_correlated_2021,
    title = {Correlated electron-hole state in twisted double-bilayer graphene},
    volume = {373},
    issn = {0036-8075, 1095-9203},
    url = {https://www.science.org/doi/10.1126/science.abc3534},
    doi = {10.1126/science.abc3534},
    number = {6560},
    urldate = {2022-09-22},
    journal = {Science},
    author = {Rickhaus, Peter and de Vries, Folkert K. and Zhu, Jihang and Portoles, Elías and Zheng, Giulia and Masseroni, Michele and Kurzmann, Annika and Taniguchi, Takashi and Watanabe, Kenji and MacDonald, Allan H. and Ihn, Thomas and Ensslin, Klaus},
    month = sep,
    year = {2021},
    pages = {1257--1260},
}

@article{rickhaus_gap_2019,
    title = {Gap {Opening} in {Twisted} {Double} {Bilayer} {Graphene} by {Crystal} {Fields}},
    volume = {19},
    issn = {1530-6984},
    url = {https://doi.org/10.1021/acs.nanolett.9b03660},
    doi = {10.1021/acs.nanolett.9b03660},
    number = {12},
    urldate = {2025-07-20},
    journal = {Nano Letters},
    publisher = {American Chemical Society},
    author = {Rickhaus, Peter and Zheng, Giulia and Lado, Jose L. and Lee, Yongjin and Kurzmann, Annika and Eich, Marius and Pisoni, Riccardo and Tong, Chuyao and Garreis, Rebekka and Gold, Carolin and Masseroni, Michele and Taniguchi, Takashi and Wantanabe, Kenji and Ihn, Thomas and Ensslin, Klaus},
    month = dec,
    year = {2019},
    pages = {8821--8828},
}

@article{scarola_phase_2001,
    title = {Phase diagram of bilayer composite fermion states},
    volume = {64},
    url = {https://link.aps.org/doi/10.1103/PhysRevB.64.085313},
    doi = {10.1103/PhysRevB.64.085313},
    number = {8},
    urldate = {2026-04-28},
    journal = {Physical Review B},
    publisher = {American Physical Society},
    author = {Scarola, V. W. and Jain, J. K.},
    month = aug,
    year = {2001},
    pages = {085313},
}

@article{alicea_interlayer_2009,
    title = {Interlayer {Coherent} {Composite} {Fermi} {Liquid} {Phase} in {Quantum} {Hall} {Bilayers}},
    volume = {103},
    url = {https://link.aps.org/doi/10.1103/PhysRevLett.103.256403},
    doi = {10.1103/PhysRevLett.103.256403},
    number = {25},
    urldate = {2026-04-28},
    journal = {Physical Review Letters},
    publisher = {American Physical Society},
    author = {Alicea, Jason and Motrunich, Olexei I. and Refael, G. and Fisher, Matthew P. A.},
    month = dec,
    year = {2009},
    pages = {256403},
}

@article{geraedts_competing_2015,
    title = {Competing {Abelian} and non-{Abelian} topological orders in $\nu=1/3+1/3$ quantum {Hall} bilayers},
    volume = {91},
    url = {https://link.aps.org/doi/10.1103/PhysRevB.91.205139},
    doi = {10.1103/PhysRevB.91.205139},
    number = {20},
    urldate = {2026-04-28},
    journal = {Physical Review B},
    publisher = {American Physical Society},
    author = {Geraedts, Scott and Zaletel, Michael P. and Papić, Zlatko and Mong, Roger S. K.},
    month = may,
    year = {2015},
    pages = {205139},
}

@article{li_excitonic_2017,
    title = {Excitonic superfluid phase in double bilayer graphene},
    volume = {13},
    copyright = {2017 Springer Nature Limited},
    issn = {1745-2481},
    url = {https://www.nature.com/articles/nphys4140},
    doi = {10.1038/nphys4140},
    number = {8},
    urldate = {2026-01-05},
    journal = {Nature Physics},
    publisher = {Nature Publishing Group},
    author = {Li, J. I. A. and Taniguchi, T. and Watanabe, K. and Hone, J. and Dean, C. R.},
    month = aug,
    year = {2017},
    pages = {751--755},
}

@misc{nguyen_bilayer_2025,
    title = {Bilayer {Excitons} in the {Laughlin} {Fractional} {Quantum} {Hall} {State}},
    url = {http://arxiv.org/abs/2410.24208},
    doi = {10.48550/arXiv.2410.24208},
    urldate = {2025-10-01},
    publisher = {arXiv},
    author = {Nguyen, Ron Q. and Zhang, Naiyuan J. and Khurana-Batra, Navketan and Liu, Xiaoxue and Watanabe, Kenji and Taniguchi, Takashi and Feldman, D. E. and Li, J. I. A.},
    month = jul,
    year = {2025},
    note = {arXiv:2410.24208 [cond-mat]},
}

@misc{han_anyon_2025,
    title = {Anyon superfluidity of excitons in quantum {Hall} bilayers},
    url = {http://arxiv.org/abs/2508.14894},
    doi = {10.48550/arXiv.2508.14894},
    urldate = {2025-08-21},
    publisher = {arXiv},
    author = {Han, Zhaoyu and Wang, Taige and Dong, Zhihuan and Zaletel, Michael P. and Vishwanath, Ashvin},
    month = aug,
    year = {2025},
    note = {arXiv:2508.14894 [cond-mat]},
}

@article{liu_interlayer_2019,
    title = {Interlayer fractional quantum {Hall} effect in a coupled graphene double layer},
    volume = {15},
    copyright = {2019 The Author(s), under exclusive licence to Springer Nature Limited},
    issn = {1745-2481},
    url = {https://www.nature.com/articles/s41567-019-0546-0},
    doi = {10.1038/s41567-019-0546-0},
    number = {9},
    urldate = {2026-01-06},
    journal = {Nature Physics},
    publisher = {Nature Publishing Group},
    author = {Liu, Xiaomeng and Hao, Zeyu and Watanabe, Kenji and Taniguchi, Takashi and Halperin, Bertrand I. and Kim, Philip},
    month = sep,
    year = {2019},
    pages = {893--897},
}


\noindent \textbf{Acknowledgements}

We thank Gil Young Cho for helpful discussions. We acknowledge the support of SNSF (Eccellenza Grant No. PCEGP2\_194528) and support from the QuantERA II Programme, which has received funding from the European Union’s Horizon 2020 research and innovation programme (Grant Agreement No. 101017733). K.W. and T.T. acknowledge support from the JSPS (KAKENHI Grant Nos. 20H00354 and 23H02052) and World Premier International Research Center Initiative, MEXT, Japan. G.W. is supported by the SNSF (Ambizione Grant No. PZ00P2-216183). 

\noindent \textbf{Author Contribution}

M.B. and N.M. conceived the project. N.M. made the stacks and fabricated the devices and performed the measurements. N.M. has analyzed the data with inputs from M.B. K.W. and T.T. provided the hBN crystals. N.M. and M.B. wrote the manuscript.

\noindent \textbf{Competing Interests}

\noindent The authors declare that they have no competing interests.

\newpage




\end{document}